\documentclass[9pt,twocolumn,twoside]{osajnl}

\journal{ol} 

\setboolean{shortarticle}{true}

\title{Terahertz vortex wave generation in air plasma by two-color femtosecond laser pulses}

\author[1,*]{Maksym Ivanov}
\author[2]{Illia Thiele}
\author[3]{Luc Berg\'e}
\author[4]{Stefan Skupin}
\author[1]{Danas Buo\v{z}ius}
\author[1]{Virgilijus Vai\v{c}aitis}

\affil[1]{Laser Research Center, Vilnius University, 10 Sauletekio avenue, Vilnius, Lithuania, LT-10223}
\affil[2]{Department of Physics, Chalmers University of Technology, G{\"o}teborg, Sweden, SE-412 96}
\affil[3]{CEA, DAM, Arpajon, France, DIF - 91297}
\affil[4]{Institut Lumiere Matiere, UMR 5306 Université Lyon 1 - CNRS, Université de Lyon, Villeurbanne, France, F-69622}

\affil[*]{Corresponding author: maks.ivannov@gmail.com}

\begin{abstract}
We investigate the generation of broadband terahertz (THz) pulses with phase singularity from air plasmas created by fundamental and second harmonic laser pulses. We show that when the second harmonic beam carries a vortex charge, the THz beam acquires a vortex structure as well. A generic feature of such THz vortex is that the intensity is modulated along the azimuthal angle, which can be attributed to the spatially varying relative phase difference between the two pump harmonics. Fully space and time resolved numerical simulations reveal that transverse instabilities of the pump further affect the emitted THz field along nonlinear propagation, which produces additional singularities resulting in a rich vortex structure. The predicted intensity modulation is experimentally demonstrated with a thermal camera, in excellent agreement with simulation results. The presence of phase singularities in the experiment is revealed by astigmatic transformation of the beam using a cylindrical mirror.
\end{abstract}

\setboolean{displaycopyright}{true}

\begin{document}

\maketitle

\section{Introduction}
Terahertz (THz) radiation is of great current interest due to many applications such as nonlinear THz spectroscopy and imaging~\cite{2011Jepsen} or electron bunch compression~\cite{2016Kealhofer}. A compact and effective method to obtain very high THz field strengths and extremely broadband spectral widths is THz wave generation from plasma filaments formed in air by focused bichromatic femtosecond laser pulses consisting of first harmonic (FH) and second harmonic (SH) waves \cite{2000Cook, 2012Clough,Thomson:lpr:1:349,2010Babushkin}. On the other hand, special light fields, such as optical vortex \cite{2016Ji}, radially polarized \cite{2017Stanislovaitis}, Bessel \cite{1987Durnin, 2003Vaicaitis} and Airy beams \cite{2007Siviloglou} are widely studied and employed in various fields \cite{1994Hell,2018Ivanov}. 
Previous attempts on vortex generation at THz frequencies \cite{2013He, 2014Imai, 2016Miyamoto, 2017Ge, 2017GeCrystals, 2017Minasyan, 2017Liu, 2018Wu} were exclusively based on manipulation of THz waves by employing external components which are inherently limited in terms of acceptable bandwidth. Therefore, alternative methods for ultra-broadband vortex generation at THz frequencies should be proposed and investigated. Very recently, vortex-shaped THz pulses have been already generated without external shaping elements \cite{2019Lin, 2019Dhaybi} in ZnTe crystal, which, however, supply relatively narrow bandwidths. 

In this work we investigate vortex THz pulse generation in an air-plasma induced by the coupling between Gaussian FH and vortex SH pulses. In this novel scheme, the vorticity is created already at the THz generation stage.
It appears that SH carrying an optical vortex charge affects not only the phase but also the intensity distribution of the generated THz pulse. 
We distinguish two stages of THz vortex generation: (i) At the beginning of the plasma filament, an intensity modulated THz vortex pulse is created. This intensity modulation is frequency dependent. (ii) Upon further propagation, the pump pulse may undergo spatio-temporal instabilities which induce secondary phase singularities in the THz field, but the total topological charge is conserved. 
Results of our investigation suggest an alternative method for the generation of structured THz waves spanning ultrabroadband frequency ranges.

\section{Theory}
\label{sec:theory}

\begin{figure}
    \centering
    \includegraphics[width=0.9\columnwidth]{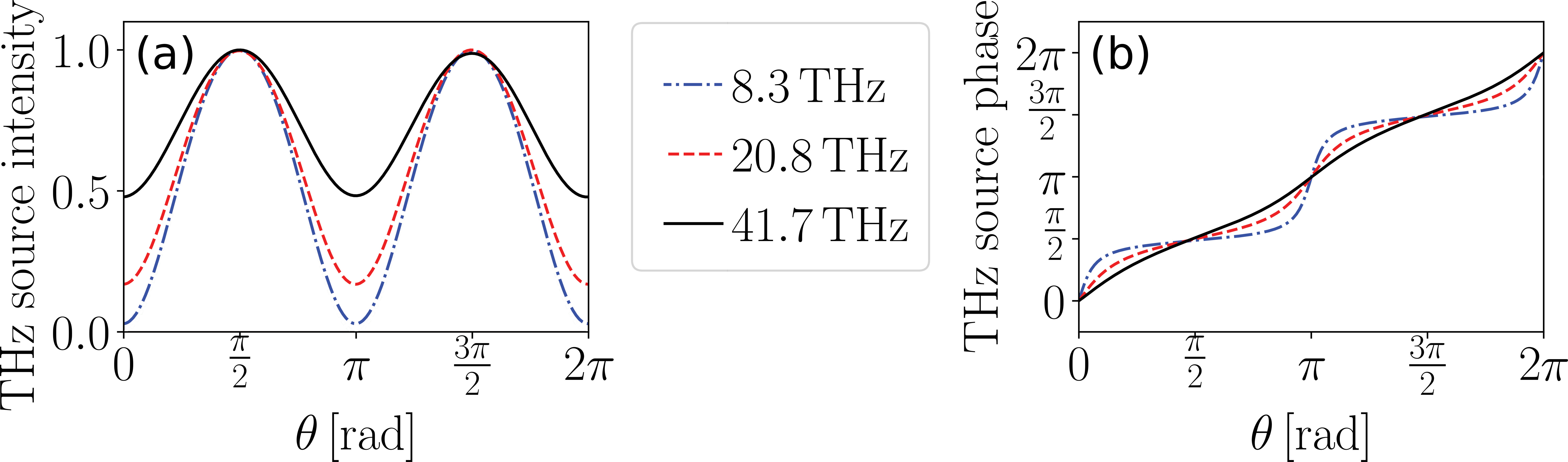}
    \caption{Azimuthal intensity~(a) and phase~(b) distribution of the Fourier-transformed THz source term $\partial_t J$ [see Eqs.~(\ref{eq:current}), (\ref{eq:E})]. 
    }
    \label{fig:Fig1_Theoretical_int_phase}
\end{figure}

For two color-laser-induced gas plasmas, the ionization current mechanism~\cite{Kim2008} is the key player for THz emission. This emission is caused by the macroscopic current of free electrons which are created by field ionization in the tunneling regime \cite{Ammosov-1986-Tunnel,PhysRevA.64.013409} and driven by the laser electric field. The principal electric field component is transverse to the laser propagation direction, and we consider linear polarization. Neglecting electron collisions, the current equation reads
\begin{equation}
    \partial_t J = \frac{q_\mathrm{e}^2}{m_\mathrm{e}} n_\mathrm{e} E\label{eq:current}\,\mathrm{,}
\end{equation}
with electron charge $q_\mathrm{e}$, mass $m_\mathrm{e}$, electron density $n_\mathrm{e}$ and electric field $E$. The electron density is a time-dependent parameter governed by the ionization rate and can be computed using ionization rate equations~\cite{Thomson:lpr:1:349,Thiele:18}. The electric field consists of both optical frequency components and generated THz components, which plays an important role in the THz spectral broadening~\cite{CabreraGranado2015}. 
However, certain effects can be already understood by considering only the impact of the bichromatic laser electric field.  The down-conversion from laser to THz frequencies takes place because of the nonlinear product between the electron density and the laser electric field. 
It can be shown that at least two laser colors, here FH and SH with a relative phase angle of $\pi/2$, are required to obtain an efficient down-conversion towards THz frequencies~\cite{Kim2007,2010Babushkin}. 
In order to understand this down-conversion process in the case of vortex pump pulses, we first evaluate Eq.~(\ref{eq:current}) for the real-valued laser electric field 
\begin{equation}
	E_{\mathrm{L}}(t,\theta) = A(t)[E_{\omega_\mathrm{L}}\cos\left(\omega_\mathrm{L}t\right)+E_{2\omega_\mathrm{L}}\cos\left(2\omega_\mathrm{L}t+\phi+l_\mathrm{SH}\theta\right)]\mbox{,}
	\label{eq:E}
\end{equation} 
with fundamental frequency $\omega_\mathrm{L}$, relative phase offset $\phi$, pulse envelope $A(t)$; $l_{\rm SH}$ is the vortex charge of the SH beam. Here, we omit the radial and longitudinal coordinates, and write the electric field as a function of time $t$ and azimuthal angle $\theta$ only. Equation (\ref{eq:E}) thus represents the rapid time- and $\theta$-dependent distribution of the electric field along the vortex ring, assuming constant FH and SH amplitudes $E_{\omega_\mathrm{L},2\omega_\mathrm{L}}$. The resulting spectral intensity and phase of the source term $\partial_t J$ obtained by Fourier transform is presented in Fig.~\ref{fig:Fig1_Theoretical_int_phase} for $\phi=0$ and $l_{\rm SH}=1$. The intensity is modulated along $\theta$ and is maximal when the relative phase angle $(\phi+l_\mathrm{SH}\theta)$ between the SH and FH takes the values $\pi/2$ and $3\pi/2$ and minimal for $0$ and $\pi$ (Fig. \ref{fig:Fig1_Theoretical_int_phase}). The modulation depth is the largest at lower frequencies and decreases for larger frequencies, where also the phase approaches a linear ramp-up along $\theta$ as expected for a vortex.
The results from this simple theoretical approach remain qualitatively the same when changing the laser and gas parameters within the parameter range allowing for efficient THz generation. 

\section{Experiments}

\begin{figure}
    \centering
    \includegraphics[width=0.9\columnwidth]{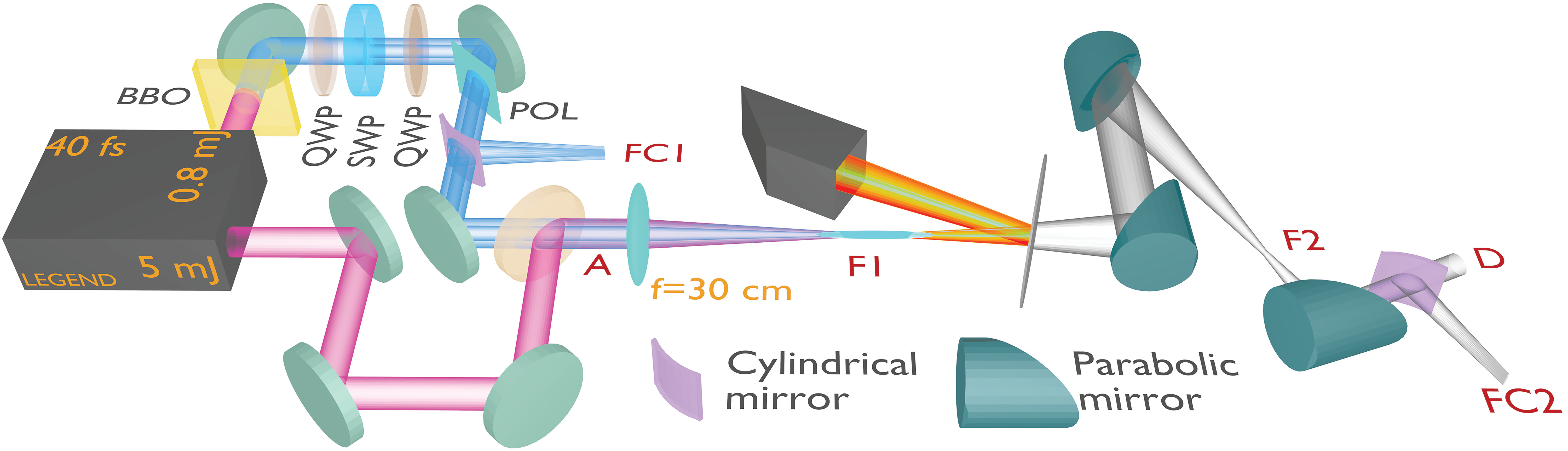}
    \caption{Sketch of experimental setup. 
    QWP - quarter wave plate, SWP - S-waveplate (q-plate), POL - polarizer, BBO - nonlinear crystal for SH generation. 
    SH is shown by blue, FH by red-pink, and THz radiation by gray color. 
    Red letters label relevant positions (see text for details). THz fluency profiles taken in positions `D' and `FC2' are shown in Fig. \ref{fig:Fig1} and \ref{fig:Cyl_focus}.}
    \label{fig:Fig2_setup}
\end{figure}

The experimental setup is sketched in Fig. \ref{fig:Fig2_setup}. We used a 1 kHz repetition rate laser system (Legend elite duo HE+, Coherent Inc.). Right before the main focusing lens (silica, $f$=30 cm) labeled `A' in the setup, the FH pulse had central wavelength of 790~nm, FWHM pulse duration 50~fs and energy of 6~mJ. The SH pulse had central wavelength of 395~nm, FWHM duration $\sim$50~fs and energy of 60~$\mu$J in Gaussian state and 50~$\mu$J in vortex state. Our Gaussian SH pulse was shaped into an optical vortex by the method described in \cite{2018Gecevicius}, which ensures more than one-octave spectral bandwidth of vortex generation: First SH was circularly polarized by a quarter wave plate (QWP) and converted to an optical vortex (OV) beam by an S-waveplate (RPC-405-06-557, Workshop of Photonics) (SWP). Subsequent polarization filtering by a second quarter wave plate (QWP) and a polarizer (POL) ensured generation of a linearly polarized vortex in the SH beam profile over the whole spectral bandwidth of the SH pulse. A 0.5 mm thick Si wafer and various commercial THz filters were used to remove the high frequency part of the pump and transmit only THz radiation, which was then collimated and shrank by parabolic mirrors in a telescope configuration to match the detector aperture. Imaging of the generated THz beam was performed with a thermal camera detector (VarioCAM head HiRes 640, InfraTec GmbH), sensitive in the range 0.1 – 40 THz (3000 – 7.5 $\mu$m). Spectra of THz radiation were obtained from Fourier transformed interferometric measurements using pyroelectric detector (TPR-A-65 THz, Spectrum Detector Inc.), sensitive in the range 0.1 - 300 THz (3000 - 1 $\mu$m) with a flat response function from $\sim$3 to $\sim$100 THz. The laser-to-THz conversion efficiency for the regular Gaussian pulses was about $10^{-4}$, but dropped to $\sim 10^{-5}$ in the case of the SH vortex pump, which we attribute to the differences in the spatial intensity distribution of the Gaussian FH and vortex SH beam.

\begin{figure}
    \centering
    \includegraphics[width=0.9\columnwidth]{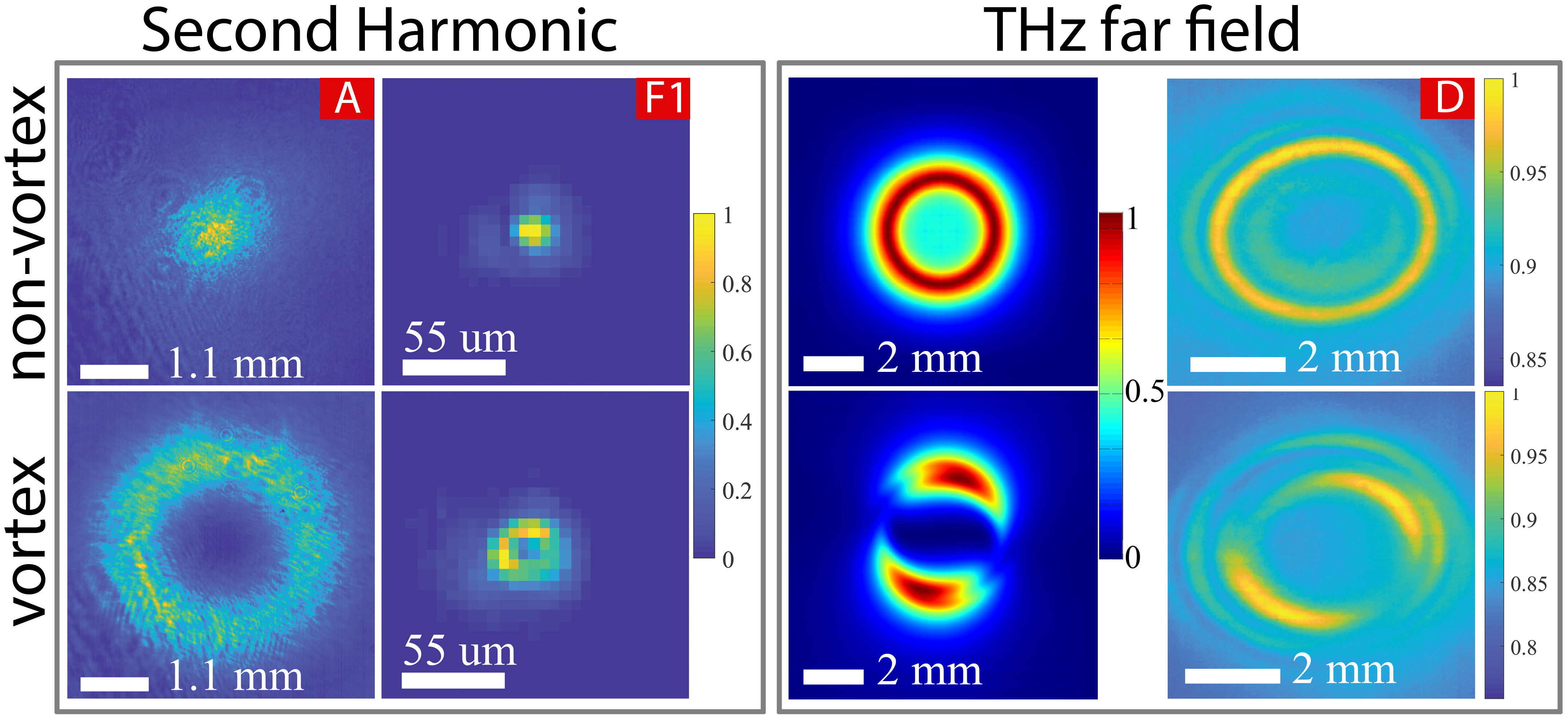}
    \caption{(Left panel) Images of second harmonic fluency (non-vortex and vortex, for Gaussian and vortex state, respectively). Column labeling correspond to locations where images were acquired as specified in Fig. \ref{fig:Fig2_setup}. (Right panel) Corresponding simulated (first column) and experimentally observed (second column) farfield THz fluencies. See text for details.}
    \label{fig:Fig1}
\end{figure}

During the experiment the FH beam was always kept Gaussian. Experimental images of the SH beam are shown in the left panel of Fig.~\ref{fig:Fig1}. The first and the second row present the SH Gaussian ('non-vortex') and vortex beam ('vortex'), respectively. 
The first column `A' shows images of the harmonics prior to the main focusing lens at position A (see Fig.~\ref{fig:Fig2_setup}); the second column `F1' refers to the focus position (F1 in Fig.~\ref{fig:Fig2_setup}). 
The FWHM beam widths are 33 and 38 $\mu$m for the FH and the vortex SH in the focal plane of the 30-cm focusing lens, respectively. 

Experimental images of the THz farfield fluency obtained in position `D' (Fig.~\ref{fig:Fig2_setup}) with the thermal camera (0.1-40 THz) are shown in the last column of Fig.~\ref{fig:Fig1}).
The THz fluency obtained with Gaussian SH (upper row) has a symmetric ring intensity distribution, as expected for conical THz emission \cite{2013Klarskov, 2018Vaicaitis}. In contrast, the THz fluency obtained with vortex SH (lower row) has an intensity modulation manifesting as two maxima along the azimuthal angle, in agreement with our theoretical predictions (see Fig. \ref{fig:Fig1_Theoretical_int_phase}). Simulation results shown for comparison in the first column of right-hand panel in Fig.~\ref{fig:Fig1}) are in excellent agreement. The simulated fluency is obtained for the spectral range of 0.01 to 50 THz (see below for details on the simulations).

\begin{figure}
    \centering
    \includegraphics[width=0.9\columnwidth]{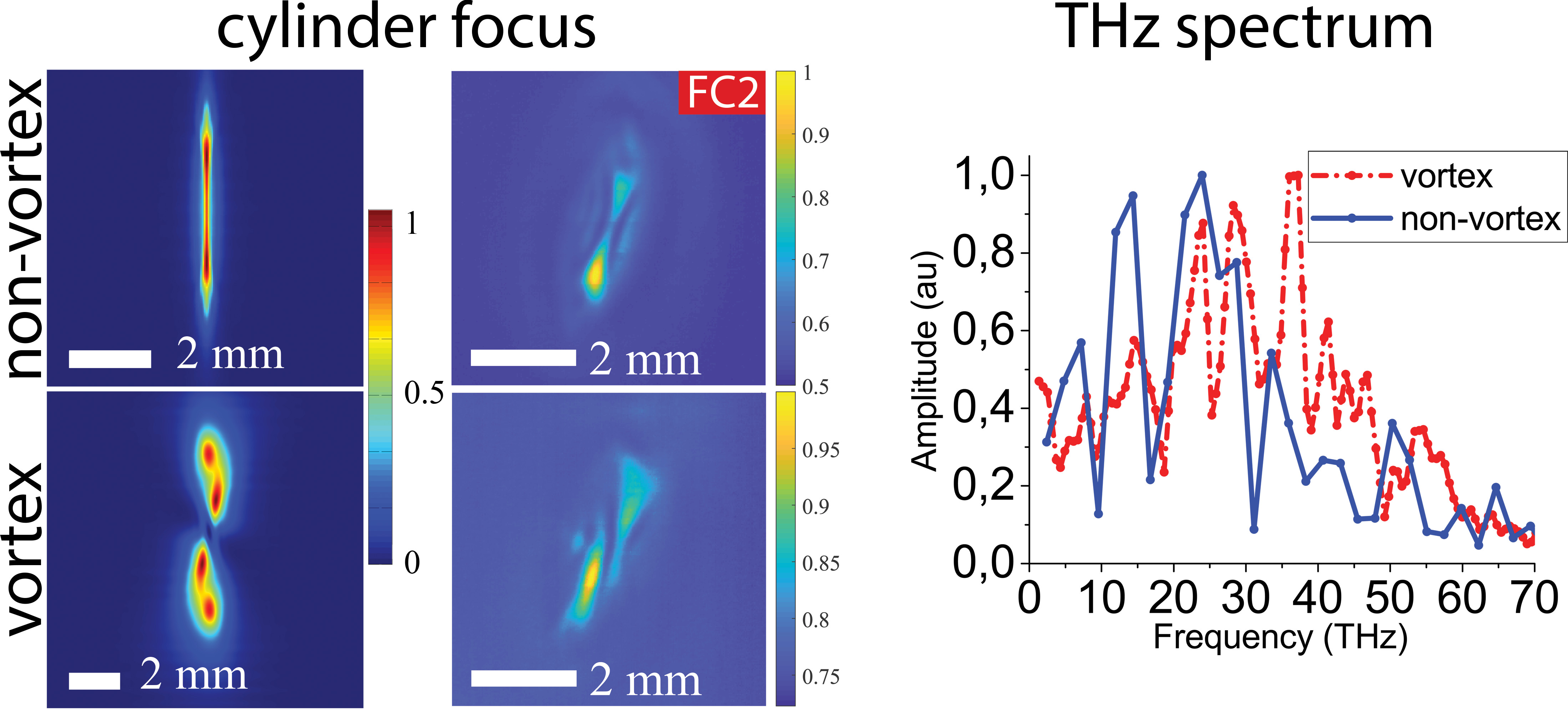}
    \caption{(Left panel) Simulated (first column) and experimentally observed (second column) fluencies of the generated THz beams in the focus of a cylindrical mirror (f~=~20~cm). The right-hand side panel shows typical experimental amplitude spectra of the generated THz pulses in the case of vortex SH (red dash-dot "vortex") and Gaussian SH (blue solid line "non-vortex").}
    \label{fig:Cyl_focus}
\end{figure}

The presence of a phase singularity in the generated THz beam was confirmed by the well-controlled method of topological charge determination based on the astigmatic transformation of singular beams by a cylindrical mirror \cite{2009Denisenko}. This method is valid even in the presence of intensity modulations affecting the generated THz beam profile. A beam without phase singularity focuses in a single line, whereas the appearance of a dark stripe in this line indicates a singular phase. The number of lines and their tilt correspond to the value and sign of the topological charge. As shown in the left panel of Fig. \ref{fig:Cyl_focus}, the single dark stripe in the second row suggests that the topological charge of the generated THz vortex is $|l_{THz}|$~=~1. Simulation results and experimental images show again excellent agreement. 

As evidenced by the right-hand panel of Fig.~\ref{fig:Cyl_focus}, the experimental spectra of the generated THz pulses are broadband, spanning from 10 to 40~THz. We attribute the dips in these spectra to absorbance in optical elements such as Si filters and pellicle beam splitters. In the case of the THz vortex the peak spectral intensity is located around 35 THz, while THz pulses without phase singularity have their maximum around 25~THz. This shift may occur because of the lower ionization rate at the periphery of the FH pump beam where the SH vortex has maximum intensity and due to transverse phase variations. 

An astigmatic transformation of the generated vortex THz beam by a cylindrical mirror was used as the primary experimental detection method of the phase singularity, yielding patterns in excellent agreement with our simulation results. However, we also looked at alternative methods to confirm that our understanding of intensity modulated THz vortex production is indeed correct.
According to our theory, the azimuthal intensity modulation comes from the fact that the phase of the SH vortex beam changes by 2$\pi$ over the full azimuthal angle, while the phase of the FH is constant along this angle. Thus, changing the relative phase between FH Gaussian and SH vortex should result in a rotation of the THz vortex intensity modulation. 
To verify this property, the THz vortex intensity distribution was recorded by the thermal camera over several minutes (not shown). 
The relative phase between FH and SH was not under control and fluctuated due to long beam pass distances ($>5$~meters for each harmonic), vibrations in the room/building, temperature variations, etc. Because of the random nature of the fluctuations, the THz vortex intensity distribution clearly jittered and made even a full rotation once in a while. In contrast, when we used FH and SH Gaussian beam profiles, the intensity distribution of the THz beam (without phase singularity) did not change.
The maximum of the THz beam just slightly oscillated around the same position due to (flaws in) the pointing stability of the laser. 

\section{Simulations}

\begin{figure}
    \centering
    \includegraphics[width=0.9\columnwidth]{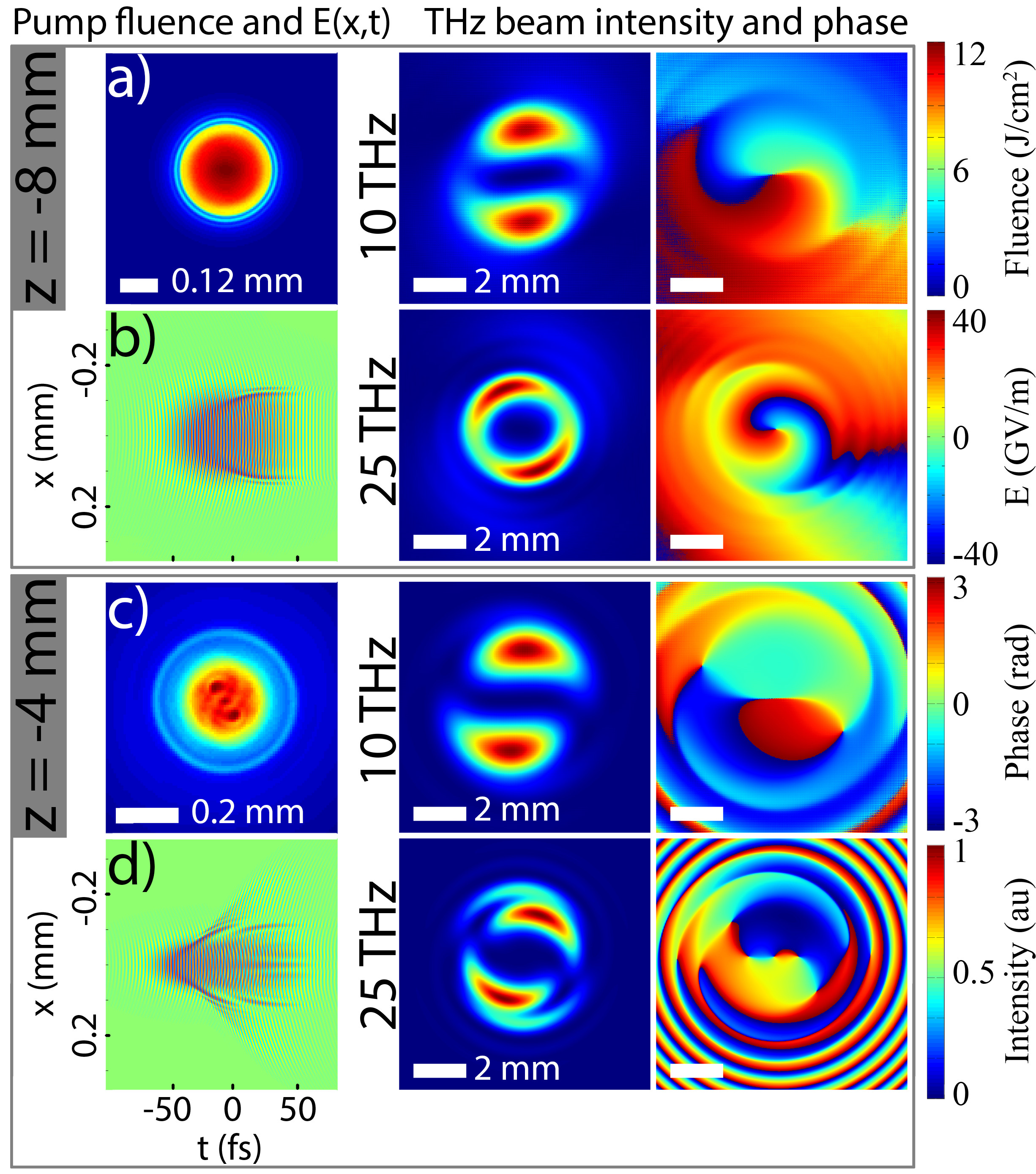}
    \caption{(a) and (c) show fluency, (b) and (d) show electric field $E(x,t)$ of the two-color pump at the beginning of the filament ($z=-8$~mm) and after transverse pump instabilities have fully developed ($z=-4$~mm). The two right-hand side columns show collimated far field intensity and phase of the 10 and 25 THz frequency components computed from the field at $z=-8$~mm and $z=-4$~mm, respectively (see text for details). Linear focus F1 is at $z=0$ (see Fig.~\ref{fig:Fig2_setup}).} 
    \label{fig:along_z}
\end{figure}

For a complementary analysis we performed comprehensive numerical simulations of the full experiment by means of a unidirectional pulse propagation solver~\cite{Kolesik:pre:70:036604,2010Babushkin}. Laser parameters and focusing conditions in the simulations are chosen such as in the experiments. Our simulation results reveal two distinct stages in the formation and subsequent evolution of the generated THz pulses. First, intensity and phase modulations of the THz source term due to the spatially variable relative phase difference between FH and vortex SH pump components emerge, as expected from our theoretical predictions (see Fig. \ref{fig:Fig1_Theoretical_int_phase}). 
Second, the pump pulse develops spatio-temporal instabilities generic to the filamentation dynamics that directly affect the THz pulse distribution.  

The right hand side of Fig.~\ref{fig:along_z} shows collimated far field intensities and phases of the generated 10 and 25 THz spectral components at two stages of the propagation, namely at the beginning of the plasma filament ($z=-8$~mm), and after transverse pump instabilities have fully developed ($z=-4$~mm).
All $z$ positions are given relative to the focus position F1 ($z = 0$ mm) of the main pump focusing lens (see Fig.~\ref{fig:Fig2_setup}).
To obtain the THz far field, the nonlinear interaction was stopped at the given $z$ position and the THz field was further propagated over a few centimeters in vacuum (sufficiently to reach the far field) and collimated with a lens in thin element approximation (quadratic phase).

At the beginning of the filament ($z=-8$~mm) the pump fluency is uniform, as evidenced by Fig.~\ref{fig:along_z}~(a). The corresponding spatio-temporal electric field profile in Fig.~\ref{fig:along_z}~(b) shows that plasma has only started to deplete the trailing part of the pump pulse, as one would expect from the usual filamentation dynamics~\cite{Kolesik:pre:70:036604,Luc:05}.
The generated THz vortex at this early stage of pump propagation has a single phase singularity in the center. The intensity is circular with azimuthal modulation, and the modulation depth is larger for lower frequencies, as predicted by Fig.~\ref{fig:Fig1_Theoretical_int_phase}. In contrast, at  ($z=-4$~mm) the pump pulse has undergone severe transverse distortion [see fluency in Fig.~\ref{fig:along_z}~(c)], and the spatio-temporal field profile in Fig.~\ref{fig:along_z}~(d) looks much more complex. These strong perturbations of the pump clearly affect the intensity distribution of the THz vortex and produce secondary singularities in the generated THz field. Nevertheless, additional singularities have alternating signs such that the total topological charge of the singular THz $|l_{THz}|=1$ is preserved during propagation at all relevant frequency components. Overall, our simulations results suggest that the produced THz vortices are surprisingly stable against pump distortions.


\section{Conclusions}

We have investigated the properties of THz radiation generated in air plasma by focused bichromatic femtosecond laser pulses, when one of the pump beams (second harmonic) is an optical vortex.
The presence of a phase singularity in the generated THz beam was confirmed by astigmatic transformation of the singular THz beams in the focus of a cylindrical lens, as well as by fully space and time resolved numerical simulations. 
We report that, in contrast to other nonlinear processes (second harmonic generation, parametric generation, etc.), the THz radiation generated by electron currents in a plasma filament can not be characterized as a THz vortex beam in the `classical' sense, such as a pure Laguerre-Gaussian beam. 
Instead, the intensity of the THz beam is modulated along the beam azimuthal angle and contains two minima between two lobes of maximum intensity. 
This is because the relative phase between two harmonics varies azimuthally when the SH pump pulse is a vortex. Moreover, our numerical simulations demonstrate that transverse instabilities in the filamentary pump propagation affect the THz vortex without destroying it. They may introduce secondary phase singularities, which renders the phase topology of produced structured THz fields particularly rich. One of the benefits of THz generation from plasma currents is the large ($>40$~THz) spectral range achievable, contrary to bandwidth limited external THz shaping techniques. 
We envisage that different combinations of the topological charges of the FH and SH pulses open a wide playground for the creation of structured singular THz sources.

\bigskip

\textbf{Funding.} Laserlab-Europe EU-H2020 654148. This research was funded by a grant (No. S-MIP-19-46) from the Research Council of Lithuania.
Simulations were performed using HPC resources from Grand Équipement National De Calcul Intensif (Grant No. A0060507594).

\textbf{Acknowledgments.} Authors are grateful to Titas Gertus from Workshop of Photonics, Vilnius, Lithuania, for the S-waveplate used in this experiment as an optical vortex generator.

\bibliography{sample}

\bibliographyfullrefs{sample}


\ifthenelse{\equal{\journalref}{aop}}{%
\section*{Author Biographies}
\begingroup
\setlength\intextsep{0pt}
\begin{minipage}[t][6.3cm][t]{1.0\textwidth} 
  \begin{wrapfigure}{L}{0.25\textwidth}
    \includegraphics[width=0.25\textwidth]{john_smith.eps}
  \end{wrapfigure}
  \noindent
  {\bfseries John Smith} received his BSc (Mathematics) in 2000 from The University of Maryland. His research interests include lasers and optics.
\end{minipage}
\begin{minipage}{1.0\textwidth}
  \begin{wrapfigure}{L}{0.25\textwidth}
    \includegraphics[width=0.25\textwidth]{alice_smith.eps}
  \end{wrapfigure}
  \noindent
  {\bfseries Alice Smith} also received her BSc (Mathematics) in 2000 from The University of Maryland. Her research interests also include lasers and optics.
\end{minipage}
\endgroup
}{}

\end{document}